\documentclass[aps,prl,amsmath, amssymb, english, aps, prl, twocolumn,reprint,floatfix, superscriptaddress]{revtex4-1}

\usepackage[english]{babel}
\usepackage{graphicx}
\usepackage{verbatim}

\usepackage{epsfig}
\usepackage{epstopdf}
\usepackage{amsfonts}
\usepackage{amsthm}
\usepackage{amsmath}
\usepackage{amssymb}
\usepackage{color}
\usepackage[usenames,dvipsnames,svgnames,table]{xcolor}
\usepackage{hyperref} 
\hypersetup{backref,pdfpagemode=FullScreen,colorlinks=true,linkcolor=NavyBlue,citecolor=BrickRed}
\usepackage{makeidx}
\usepackage{pifont}
\usepackage{pdfpages}

\usepackage{color}
\usepackage{grffile}


\def\expect#1{\mathinner{\langle{#1}\rangle}}

{\catcode`\|=\active
  \gdef\expect#1{\left<\mathcode`\|"8000\let|\bravert {#1}\right>}}
\def\bravert{\egroup\,\vrule\,\bgroup}

\def\beq{\begin{equation}}
\def\eeq{\end{equation}}
\def\be{\begin{equation}}
\def\ee{\end{equation}}

\def\epsk{\epsilon_{{\bf k}}}

\def\cG0{{\cal G}_0}

\def\spinup{\uparrow}
\def\spindown{\downarrow}

\def\d{\delta}

\def\eps{\epsilon}

\def\s{\sigma}

\def\uc2{$U_{c2}$}
\def\uc1{$U_{c1}$}

\def\bea{\begin{eqnarray}}
\def\eea{\end{eqnarray}}
\def \bal{\begin{align}}
\def \eal{\end{align}} 
\def\#{\!\!}
\def\@{\!\!\!\!}

\def\+{\dagger}

\def\up{\spinup}
\def\down{\spindown}

\begin{document}

\title{\bf Hund's induced Fermi-liquid instabilities and enhanced quasiparticle interactions}
\author{Luca de'~Medici}
\affiliation{European Synchrotron Radiation Facility, 71 Av. des Martyrs, Grenoble, France}
\affiliation{Laboratoire de Physique et Etude des Mat\'eriaux, UMR8213 CNRS/ESPCI/UPMC, Paris, France}

\begin{abstract}

Hund's coupling is shown to generally favor, in a doped half-filled Mott insulator, an increase in the compressibility culminating in a Fermi-liquid instability towards phase separation.
The largest effect is found about the frontier between an ordinary and an orbitally-decoupled ("Hund's") metal.
The increased compressibility implies an enhancement of quasiparticle scattering, thus favoring other possible symmetry breakings. 
This physics is shown to happen in simulations of the 122 Fe-based superconductors, possibly implying the relevance of this mechanism in the enhancement of the the critical temperature for superconductivity.
\end{abstract}

\maketitle

A wealth of unexpected phenomena have been discovered in strongly correlated materials and many technological applications are foreseen. 
At the heart of the observed remarkable behaviors is the many-body physics of the conduction electrons. Indeed their tendency to avoid each other leads to a complex dynamics with surprising properties, even more so in the typical multi-orbital landscape of these systems. 
In particular the different repulsion that two electrons feel depending on them being in the same or a different orbital, or on the alignment of their spins (embodied by the well known Hund's rules of atomic physics) plays an important role in these materials.

Recently our understanding of the influence of Hund's coupling on the metallicity properties of correlated materials has drastically improved\cite{Georges_annrev}.
Three aspects have been singled out as the most influential. i) Hund's coupling tunes the splitting of atomic multiplets, that affects the distance in energy between the various sectors of atomic states with a given total charge. This energy is responsible for the local charge fluctuations, in a material, and thus for the ease with which a Mott insulating state can occur. The main outcome is that a Mott insulating state is strongly favored when conduction bands arise from an atomic shell which is half-filled.
ii) Hund's coupling typically lowers the overall coherence of conduction electrons, especially for electron densities near half-filling.
iii) It also favors the differentiation of the correlation strength among the conduction electrons. This was termed "orbital decoupling" \cite{demedici_MottHund,demedici-SpringerBook}, since it stems from Hund's coupling suppression of orbital fluctuations, favoring selective Mott physics depending on the orbital character of the conduction electrons. This can cause the coexistence of electrons with different correlation strength.

All of these results have been consistently found in Iron-based superconductors and related materials (see for example \cite{McNally_BaMn2As2_Mott_Hund} for i),\cite{Hardy_KFe2As2_Heavy_Fermion} for ii), \cite{demedici_OSM_FeSC,Yi_Universal_OSM_Chalcogenides} for iii)) thus testifying the importance of electronic correlations in these compounds in accord with theoretical studies\cite{Ishida_Mott_d5_nFL_Fe-SC,demedici_OSM_FeSC,Shorikov_LaFeAsO_OSMT,*Aichhorn_FeSe,*Yin_kinetic_frustration_allFeSC,*Werner_122_dynU,*Misawa_d5-proximity_magnetic,*Backes_KRbCs122,*Bascones_FeSC_Magn_Review,*Si_NatureReview}.

However these materials are Fermi-liquids at low temperature\cite{Rullier_Review} and their instabilities (magnetism, superconductivity) have been consistently modeled within weak-coupling theories\cite{Mazin_Splusminus,*Chubukov_ItinerantScenario,*Hirschfeld_GapSymmetry}.
The influence of electronic correlations on the high-T$_c$ superconductivity is at present still not understood.

Here we show that multi-orbital correlations and Hund's coupling in particular seem to have another, hitherto undiscovered, influence on these systems, in that not only the quasiparticle weight and masses but also the residual interactions between quasiparticles can be affected in a highly non-trivial way. This happens in the zone of influence of the aforementioned half-filled Mott insulator. We show that Hund's coupling affects these interactions enhancing the compressibility of the electron fluid, up to a point in which the system is unstable towards phase separation. Besides the many implications that an intrinsic instability towards a phase separation can entail, it is worth stressing that simply these altered quasiparticle interactions can have direct effects on the interaction vertices with low-energy bosons and radically enhance some bosonic-mediated mechanisms towards long-range ordered phases. We will show that this might be the case for the pairing mechanism for the high-Tc superconductivity in Fe-based superconductors.

We investigate the multi-orbital Hubbard model (with M orbitals) with Hamiltonian $\hat H-\mu \hat N=\hat H_0+\hat H_{int}-\mu \hat N$ with
\be
\hat H_0 = \sum_{i\neq jmm'\s} t^{m m'}_{ij} d^\dag_{im\s}d_{jm'\s}+\sum_{im\s}\eps_mn_{im\s},
\ee
where $d^\dag_{im\s}$ creates an electron with spin $\s$ in orbital $m=1,\ldots, M$ on site $i$ of the lattice, and 
\begin{eqnarray}
\hat H_{int}\,&&=\,U\sum_m n_{m\up}n_{m\down}\,\\
\,&&+U^\prime\sum_{m\neq m'} n_{m\up}n_{m'\down}\,
+(U^\prime-J) \sum_{m< m',\sigma} n_{m\s}n_{m'\s} + \nonumber \\
&&-J\!\!\sum_{m\neq m'} d^+_{m\spinup}d_{m\spindown}\,d^+_{m'\spindown}d_{m'\spinup} \,
\!+\! J\!\! \sum_{m\neq m'} d^+_{m\spinup}d^+_{m\spindown}\,d_{m'\spindown}d_{m'\spinup}. \nonumber
\label{eq:ham_kanamori} 
\end{eqnarray}
The number operator is $n_{im\s}=d^\dag_{im\s}d_{im\s}$, $\hat N=\sum_{im\s}n_{im\s}$, $\mu$ is the chemical potential, and customarily we set $U^\prime=U-2J$ and we drop the last two terms in $\hat H_{int}$ (spin-flip and pair-hopping), which needs extra approximations to be treated within our method of choice (the effect of these terms is addressed in the Supplemental Material). We first study the degenerate model in which we only consider diagonal hopping in orbital space, equal for all orbitals i.e.   $t^{m m'}_{ij}=t_{ij}\d_{mm^\prime}$ and $\eps_m=0$, as a function of the average density of electrons per lattice site $n=\langle \hat N \rangle/{\cal N}_{sites}$.

We treat the model in the Slave-spin mean-field approximation\cite{demedici_Vietri} (SSMF) and we focus on the to normal, non-magnetic, zero-temperature metallic phase. There the SSMF describes the metal as a Fermi-liquid by construction, yielding the following quasiparticle hamiltonian:
\be\label{eq:QP_ham}
H-\mu N= \sum_{km\s} (Z\epsk+\lambda-\mu) f^\+_{km\s}f_{km\s}
\ee
where $f^\+_{km\s}$ is che creation operator of a quasiparticle with momentum $k$, orbital (band) character m and spin $\s$ and $\epsk$ is the bare electronic dispersion relation which is the same for all the bands. In what follows the k-space structure of $\epsk$ is immaterial, and only the density of states (DOS) counts; we customarily choose a semi-circular DOS $D(\eps)\equiv\frac{2}{\pi D}\sqrt{1-(\frac{\eps}{D})^2}$ of bandwidth $W=2D$ for all bands. 
The renormalization parameters of the dispersion, $Z$ and $\lambda$, are calculated in the self-consistent SSMF scheme as an average on the auxiliary slave-spin hamiltonian\cite{demedici_Vietri} which treats explicitly the interaction term eq. (\ref{eq:ham_kanamori}). They thus embody the effect of the electronic interactions determining the quasiparticles: $Z$ is the quasiparticle weight and - in this family of methods that yield a purely local electronic self-energy - also the inverse of the mass enhancement, while $\lambda$ is a shift of the bare on-site energy. The Fermi liquid condition $n_f\equiv \sum_{km\s}\langle f^\+_{km\s}f_{km\s} \rangle=n$ is built-in.

\begin{figure}[h!]
\begin{center}
\includegraphics[width=4.2cm]{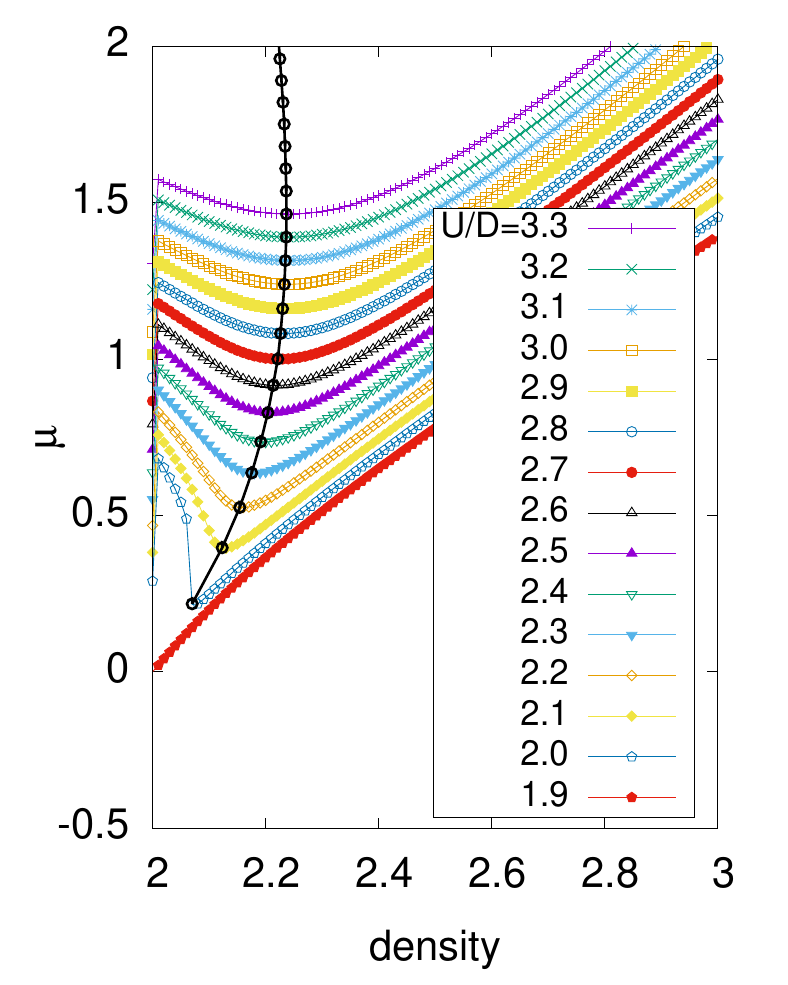}
\includegraphics[width=4.2cm]{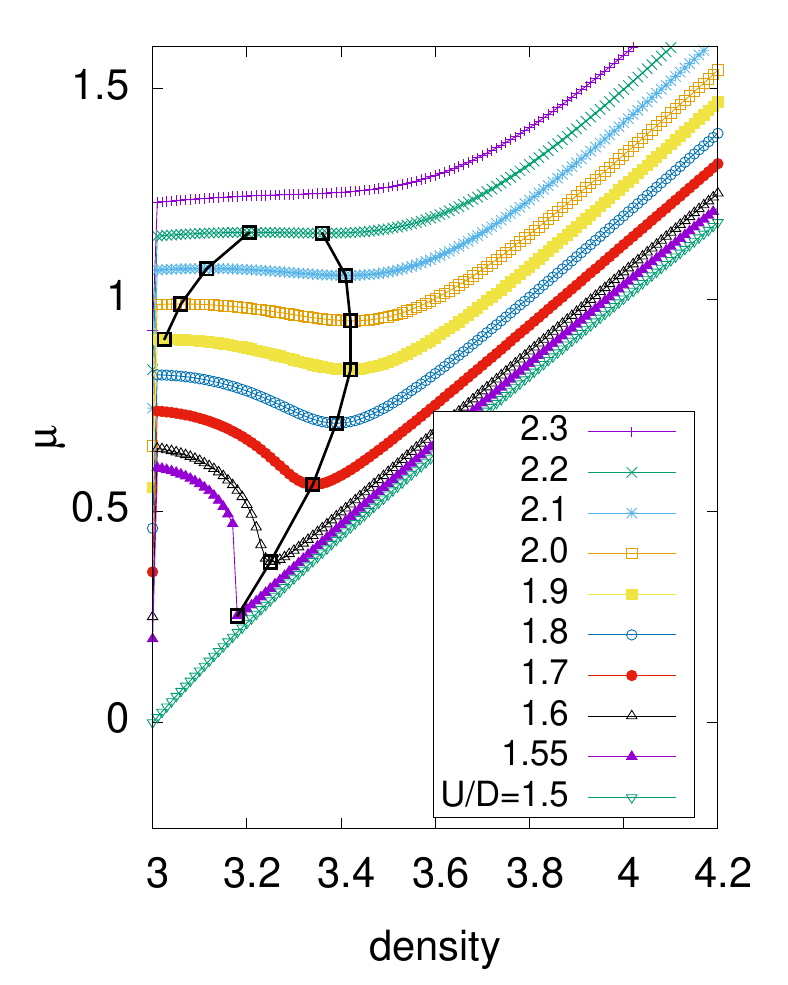}
\caption{Degenerate 2-orbital (left panel) and 3-orbital (right panel) Hubbard model with semicircular DOS of half-bandwidth D, and Ising-type Hund's coupling $J/U=0.25$: $\mu$ vs $n$ curves for different values of U. In the 2-orbital case for $U_c/D=1.96$ at half-filling ($n=2$, in this model) the system undergoes a Mott transition. The curves for $U>U_c$ show a negative slope inside a spinodal line departing from the Mott transition that is marked with black circles. In the 3-orbital case most of the $\mu$ vs $n$ curves for $U>U_c=1.515$ show a double change of slope (the same happens in the 5-orbital model, see the Supplemental Material), so that the instability zone extends between two spinodal lines (black squares) both at finite doping from half-filling (n=3, here).}
\label{fig:mu_vs_n}
\end{center}
\end{figure}

\begin{figure}[h!]
\begin{center}
\includegraphics[width=4.2cm]{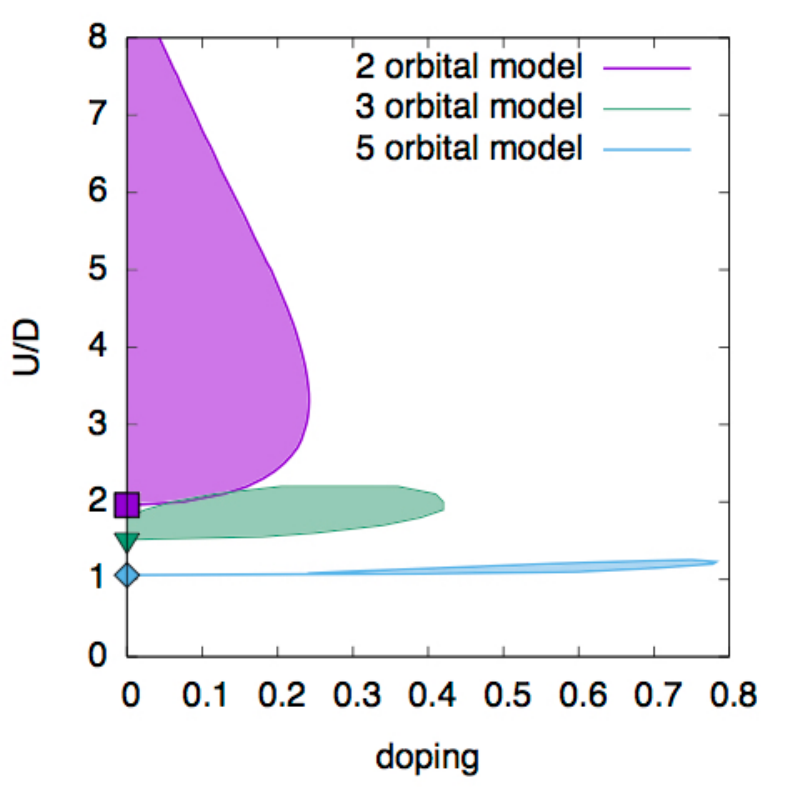}
\includegraphics[width=3.8cm]{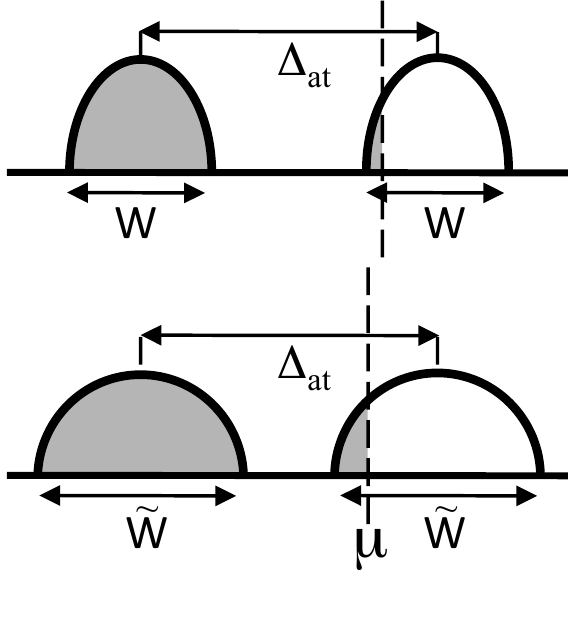}
\caption{Left panel: the colored areas are the zones of instability towards phase separation in the U-doping plane for the 2-,3-, and 5-orbital models, with $J/U=0.25$. The low-U frontier departs from the Mott transition point at half-filling (symbols). For growing number of orbitals the unstable zone extends to larger and larger doping, albeit narrowing in U. Right panel: a plausible mechanism by which Hund's coupling causes the charge instability. Hund's coupling reduces the width of the Hubbard bands at half-filling, through the quenching of orbital fluctuations. Upon doping the quenching is lifted and the Hubbard bands are expected to expand (going from $\sim W$ to some larger value $\tilde W$). This makes possible to have a lower chemical potential at larger particle density, i.e. a charge instability. }
\label{fig:PhaseDiag-Mech}
\end{center}
\end{figure}

We wish to study the compressibility of the electronic fluid $\kappa_{el}=\frac{dn}{d\mu}$ as a function of Hund's coupling and the number of orbitals in the model. We thus calculate the $\mu$ vs $n$ dependence in these models within SSMF. 
In Fig. \ref{fig:mu_vs_n} (left panel) the result for the 2-orbital Hubbard model at a typical value of Hund's coupling $J/U=0.25$ is reported for various values of U. Remarkably, approaching the half-filled Mott insulator (that is realized at $n=2$, for $U>U_c=1.96D$) the slope of the $\mu$ vs $n$ curves vanishes and then becomes negative, signaling a sizable zone in the U/doping (from half-filling, i.e. $\d\equiv n-M$) plane where the system is unstable.
Indeed a negative compressibility signals a thermodynamic instability of the system. 
Furthermore the compressibility can become negative both going through zero and through a divergence. The second case is realized here, implying a strong enhancement of the compressibility in the thermodynamically stable zone near the frontier.
The compressibility being the uniform and static charge response function, a divergence in this quantity signals an instability of the system toward phase separation into macroscopic charge-rich and charge-poor zones (which in practice will be prevented by a charge modulation at finite q due to long-range Coulomb interaction, neglected in this study\cite{CDG_PRL95}). 
This spinodal line marking the phase-separation instability in the $U-\d$ plane is shown in Fig. \ref{fig:PhaseDiag-Mech} (left panel): it departs from the Mott transition and has a non-monotonous behaviour as a function of the interaction strength, so that the unstable zone widens rapidly for $U>U_c$, but at larger interactions  (for $U/D>3.3$ in this case) it narrows again. 

Hund's coupling is essential for this instability zone to appear, no diverging compressibility is found at $J=0$.
Already at very small $J/U$ a wide zone opens at large U. This zone moves towards lower interaction strengths for increasing $J/U$, following the position of the Mott transition\cite{demedici_MottHund} at half-filling, from which the instability frontier always departs. The zone extends, for typical values of $J/U$, in the range n=2 to 2.2 (and symmetrically by respect to half-filling n=2, since the model is particle-hole symmetric), becoming larger only for $J/U\gtrsim 0.25$. The complete study as a function of $J/U$ is presented in the Supplementary Material.

A similar behavior is found for the 3-orbital (Fig. \ref{fig:mu_vs_n}, right panel) and 5-orbital model (shown in the Supplemental Material). The main difference with the 2-orbital case is however that a second frontier can be traced, at smaller doping compared to the first, where the  $\mu$ vs $n$ curve recovers a positive slope. This signals that the instability due to the compressibility divergence now mostly happens in a range of finite doping (and that the system can remain stable near the Mott insulator). The region (Fig. \ref{fig:PhaseDiag-Mech}, left panel) still departs from the Mott transition at half-filling, but then extends in the U-doping plane until the two lines merge, having thus a "moustache" shape in the $M>2$ cases, instead of an "onion" shape like in the $M=2$ case.
It is comparatively less extended in U but more extended in doping. In particular in the 5-orbital model it approaches values close to n=6 densities. At n=6 the system is stable again, but still an evident signature of this physics in terms of enhanced compressibility is present.
 
Despite a marked difference, then, in the way the instability zone evolves at large U, the common robust feature is that in all these models the onset of Hund's coupling triggers the appearance of a zone departing from the Mott transition at half-filling where the system is unstable towards phase separation. The low-U frontier of this zone in all cases is rather horizontal in the U-doping plane, i.e. for $U>U_c$ it moves quickly towards the maximum doping it will reach.
This frontier is found to follow a well known crossover, appearing in these models in several physical quantities. Indeed (see for instance Fig.16 in Ref. \cite{demedici_Vietri}) along a line departing from the Mott transition\cite{Ishida_Mott_d5_nFL_Fe-SC,Fanfarillo_Hund}, at finite doping one observes for increasing U or doping moving towards half-filling: a quick decrease of the quasiparticle weight Z, an increase of the inter-orbital spin-spin correlations, and a suppression of the inter-orbital charge correlations. The crossover is very sharp near the Mott transition, whereas it becomes progressively broader with increasing doping. At finite temperature it is identifiable with the "spin-freezing cross-over"\cite{Werner_SpinFreezing,Ishida_Mott_d5_nFL_Fe-SC,*Liebsch_FeSe_spinfreezing,*Hoshino_SC}, and the recent successful denomination of "Hund's metal"\cite{Yin_kinetic_frustration_allFeSC} might be used for the zone at large U-small doping. The three mentioned quantities then have a different, and rather disconnected evolution in other zones of the phase diagram. This crossover is where the compressibility divergence is empirically found to appear for increasing interaction U in all cases (as explicitly shown for selected cases in the Supplemental Material). The re-entrant shape of the instability zone suggests a closer analogy with the zone of reduced inter-orbital correlations, however, which is the only quantity among the three showing a similar shape in its crossover at large U \cite{demedici_Vietri}.
 
Let's now analyze the origin of this behavior. Interestingly, this phase separation arises as an instability of the Fermi-liquid, the latter being enforced, within the SSMF method. 
Indeed it is worth stressing that the renormalization parameters in the quasiparticle Hamiltonian eq. (\ref{eq:QP_ham}) depend in an intricate way on all the couplings in the bare hamiltonian and the chemical potential, so that the effective system described by eq. (\ref{eq:QP_ham}) is not to be viewed as a simple rigid band picture with an effective dispersion: the dispersion itself changes with the filling instead.
Thus besides the quasiparticle energy (the linear - in the k-resolved density -  contribution to the energy of the system) this method also accounts for the effects of quasiparticle interaction, (the quadratic term), i.e. they yield the Landau parameters $F_0^s, F_0^a$, etc\cite{nozieres}.
Indeed in an isotropic Fermi liquid metal the electronic compressibility reads:
\be\label{eq:kappa_FL}
\kappa_{el}=\frac{D^*(\eps_F)}{1+F_0^s}=\frac{1}{D^*(\eps_F)^{-1}+f_0^s}
\ee 
where $D^*(\eps_F)$ is the total quasiparticle (i.e. renormalized) density of states at the fermi energy and $F_0^s=D^*(\eps_F)f_0^s$ is the isotropic, spin-symmetric Landau parameter.  
From eq. (\ref{eq:QP_ham}) one can formally calculate the total density of quasiparticles $n_f(\mu)=\int d\eps D(\eps)n_F(Z\eps+\lambda-\mu)$ (where $n_F(\eps)$ is the Fermi function), and deriving this expression by respect to $\mu$ one indeed finds at zero temperature the above expression for the electronic compressibility with $D^*(\mu)=D(\mu_0)/Z$ and $f_0^s=\mu_0\frac{dZ}{dn}+\frac{d\lambda}{dn}$. Here $\mu_0(n)$ is the chemical potential for the non-interacting system with the same particle density ($\mu_0=0$ at half-filling in our particle-hole symmetric case), entering through the relation $\mu_0=\frac{\mu-\lambda}{Z}$ implied by the Luttinger theorem that holds in our Fermi-liquid framework (see Supplementary Material).
Since for the compressibility to diverge when Z is finite (as we indeed find here) it must be $F_0^s<-1$, and thus $f_0^s$ at least negative, this last formula implies that $\frac{d\lambda}{dn}$ has to be negative, since $\mu_0\frac{dZ}{dn}$ is always positive (indeed Z diminishes upon approaching the half-filled Mott insulator, so that its slope has always the sign of $\mu_0$). 
This is confirmed numerically (see Supplemental Material). Indeed the compressibility divergence always happens because $\frac{d\lambda}{dn}$ becomes negative and larger in absolute value than $D^*(\eps_F)^{-1}+\mu_0\frac{dZ}{dn}$. The last quantity is always small near the Mott insulator (because Z and $\mu_0$ are small there, and $\frac{dZ}{dn}$ is finite) but the question remains of why $\frac{d\lambda}{dn}<0$.

Indeed the renormalization parameters Z and $\lambda$ set respectively the width and the position of the quasiparticle band(s), compared to the non interacting case. In particular in a doped Mott insulator $\lambda$ places the quasiparticle band above the charge gap in the so-called Hubbard band\cite{Hubbard_I,*Hubbard_III}, the range of the spectrum with delocalized excitations at finite energy.
The Hubbard band is centered around the energy of the atomic charge excitation (e.g.: U/2 in the single-band case) and if its width is fixed one expects lambda to grow monotonously when shifting the quasiparticle band within it, upon doping. It has however become clear recently that the Hubbard bands can vary in width, for instance among different models: in absence of Hund's coupling their width grows like $\sqrt{M}$ with the number of orbitals\cite{Gunnarsson_fullerenes,*Gunnarsson_multiorb}, while it was shown that the onset of Hund's coupling $J$ reduces their width back to values of the order of the one-band model\cite{demedici_Vietri}. Indeed $J$ quenches the orbital fluctuations responsible for the enhancement of the delocalization energy of the charge excitations, and hence of the width of the Hubbard bands. This quenching is however complete only at half-filling, while the extra particles introduced by doping necessarily create doubly occupied orbitals, unquenching the orbital degrees of freedom. One can then expect the Hubbard bands to expand again, when doping the half-filled Mott insulator.
This gives a plausibility argument for the non-monotonic behaviour of $\lambda$ with $n$: if the Hubbard band expands quickly enough with doping it may happen that for a larger density $n$ the quasiparticle band is located at lower energy than for a smaller $n$ (see Fig. \ref{fig:PhaseDiag-Mech}, right panel), and consequently $\mu$ can be lower for the larger $n$, causing the negative compressibility that we find.

Incidentally the mechanism reducing the width of the Hubbard bands was shown to coincide\cite{demedici_Vietri} with the one causing the "orbital-decoupling" (i.e. the suppression of inter-orbital charge-charge correlations) leading to the selective Mott physics that one finds in these models once the degeneracy of the bands is removed\cite{demedici_3bandOSMT,demedici_MottHund}. It is not surprising then to find the divergence of the compressibility on the frontier of the cross-over towards the orbitally-decoupled region.

It is worth stressing that also the proximity to such a Fermi-liquid instability has several remarkable consequences\cite{CDG_PRL95}.
Indeed a negative $F_0^s$ implies an attractive interaction between quasiparticles in the particle-hole channel, at $q=0,\omega\rightarrow 0$. This in general favors superconductivity\cite{GrilliRaimondi_IntJModB}. 
But also the interaction of quasiparticles with low-energy bosons can be enhanced. Indeed Ward identities relate the quasiparticle interaction vertices with the Fermi-liquid parameters. For instance for the density-vertex $\Lambda(q,\omega)$ the following Ward identity holds\cite{Grilli_El-Ph}:
\be
Z\Lambda(q\rightarrow 0, \omega=0)=\frac{1}{1+F_0^s}.
\ee
This implies an enhancement of this interaction vertex, if $1+F_0^s$  decreases until vanishing as in the present case (for a complete plot of $1+F_0^s$ in the 2-orbital model with $J/U=0.25$ in the Supplemental Material).
In turn the enhancement of the vertex can favor a symmetry breaking, if a related susceptibility is correspondingly enhanced.

\begin{figure}[h!]
\begin{center} 
  \includegraphics[width=4.25cm]{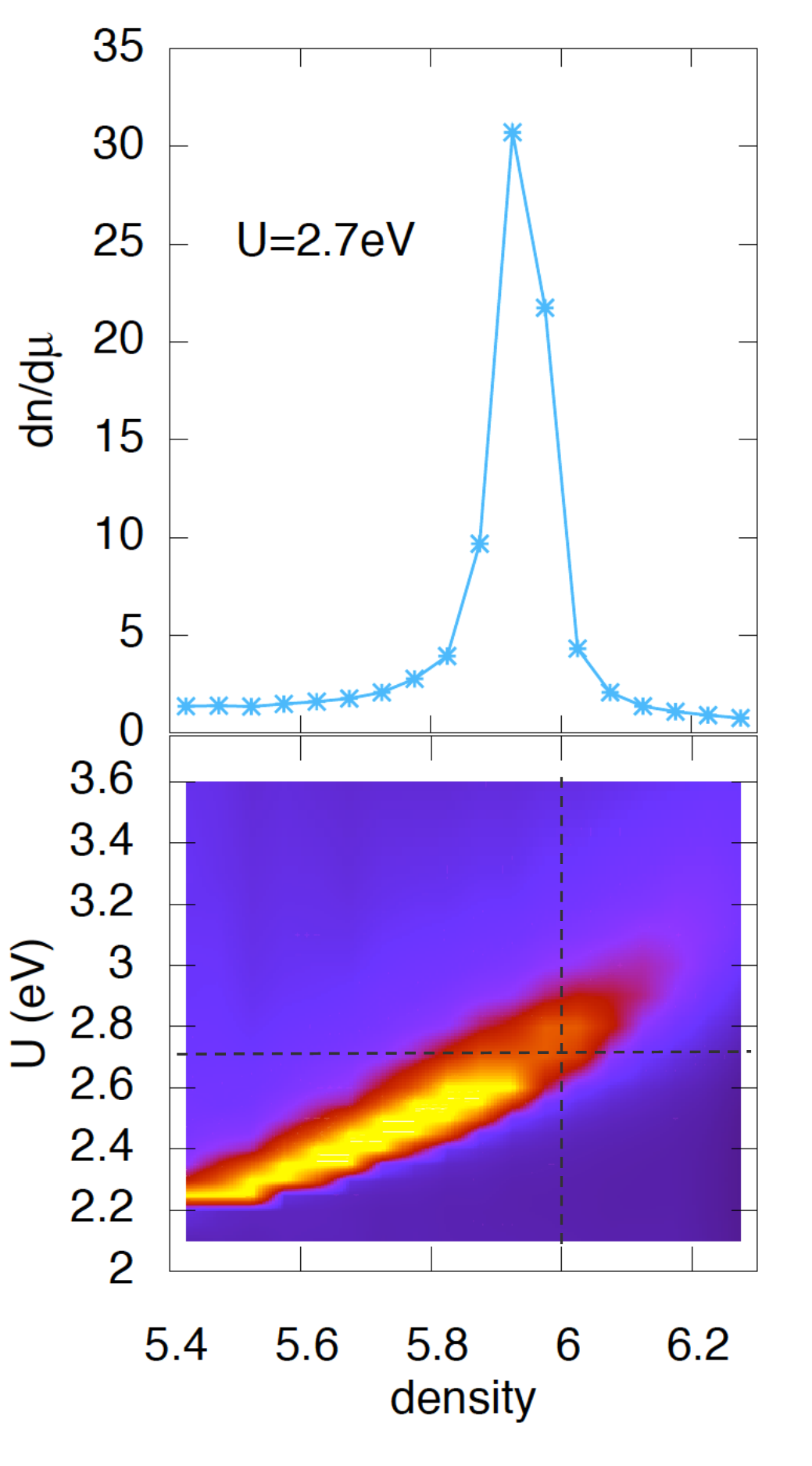}  
   \includegraphics[width=4.25cm]{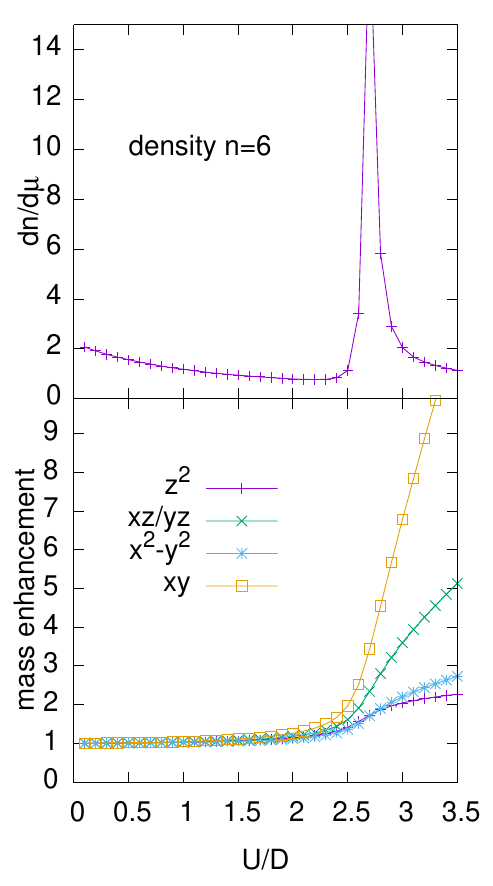}  
  \caption{Lower left panel: compressibility (color scale) in BaFe$_2$As$_2$ calculated within DFT+SSMF (J/U=0.25) in the U-density plane. The saturated yellow color corresponds to the unstable region (the pixelation is due to numerical discretization and is unphysical), surrounded by an area of enhanced compressibility (red). Upper panels: compressibility plotted along the cuts (dashed lines) for constant U=2.7eV and constant density n=6, relevant values for BaFe$_2$As$_2$. Lower right panel: orbitally resolved mass enhancements, showing that the compressibility enhancement happens about the cross-over where correlations become orbitally selective.}
  \label{fig:BaFe2As2}
\end{center}
\end{figure}
It is tempting to attribute to these effects the enhancement of a number of instabilities of the paramagnetic metallic phase in materials that are dominated by Hund's many-body physics\cite{Georges_annrev}. Most notably it is not impossible that, whatever the mechanism leading to superconductivity in Iron-based superconductors, the enhancement of the critical temperature be due to the Fermi-liquid compressibility enhancement outlined in this paper.
Indeed when modeling doped BaFe$_2$As$_2$ with DFT+Slave-spin one finds confirmation that the phase separation instability is realized also in this realistic framework, emanates from the half-filled Mott insulator, and reaches the zones in the phase diagram relevant to the iron-based superconductors, as shown in Fig. \ref{fig:BaFe2As2}.
Remarkably indeed, using the set of interaction parameters (U=2.7eV, J/U=0.25) that yields the correct Sommerfeld coefficient for all of the 122 family\cite{Hardy_122_SlaveSpin_exp,demedici_Vietri} one finds that the compressibility is greatly enhanced in the density range 5.75 to 6.1, coinciding with the zone where doped BaFe$_2$As$_2$ shows high-Tc superconductivity in the tetragonal phase.

Recently Misawa and Imada\cite{Misawa_LaFeAsO} have highlighted a zone of phase separation, in the ab-initio phase diagram of LaFeAsO   \cite{Misawa_Imada-PhaseSep_Hubbard} studied within a variational Montecarlo scheme, in proximity of the superconducting phase. The zone of phase separation is compatible with the corresponding zone we find in the present study and with its continuation in terms of enhanced compressibility. What we highlight in the present work is that this phase is a genuine instability of the Fermi liquid phase even in absence of all symmetry breaking, it is to be tracked back to Hund's coupling, and is a universal feature of all Hund's dominated doped half-filled Mott insulators.

Analogously to the model studies, we find that in BaFe$_2$As$_2$ the instability zone and the zone of enhanced compressibility that continues it, is located on the cross-over frontier between the normal and the orbitally-decoupled ("Hund's") metal (Fig. 3, right panels). We suggest that in general an enhancement of the compressibility, with the possible related enhancement of superconductive pairing, happens when a material is near the frontier between a normal metal and a Hund's metal. A possible universally detectable sign of this situation is the arising of high-Tc superconductivity in between a phase with orbitally-selective electronic correlation strength and another, more conventional metallic phase, as it also happens in Cuprates\cite{demedici_OSM_FeSC}. 

\acknowledgements The author is grateful to M. Capone, G. Giovannetti, G. Sangiovanni, M. Grilli, A. Chubukov, P. Bruno for helpful exchanges and discussions.

%

\newpage

\begin{widetext}
\begin{center}
{\bf\large Hund's induced Fermi-liquid instabilities and enhanced quasiparticle interactions}\\ 
\bigskip
{\bf \large Supplemental Material}
\end{center}
\vspace{0.5cm}
\end{widetext}

\renewcommand{\thepage}{S\arabic{page}}  
\renewcommand{\thesection}{S\arabic{section}}   
\renewcommand{\thetable}{S\arabic{table}}   
\renewcommand{\thefigure}{S\arabic{figure}}
\renewcommand{\theequation}{S\arabic{equation}}

\setcounter{page}{1}
\setcounter{figure}{0}
\setcounter{table}{0}

\section{Evolution of the zone of phase separation in the 2-orbital model as a function of Hund's coupling strength J/U}

It is interesting to study the evolution of the zone of instability towards phase separation as a function of Hund's coupling strength $J/U$. The result is shown in Fig. \ref{fig:PS_vs_ratioJU}. It is remarkable that for $J=0$ the system is always stable, while already for very small Hund's coupling a large instability zone opens up at large U. Upon increasing J, the zone moves towards lower interaction strengths, following the position of the Mott transition\cite{Sdemedici_MottHund} while becoming less extended in U: this implies that the maximum extension in density, which in this model is somewhat beyond $n=2.2$ electrons/site, i.e. the maximum distance of the frontier from the Mott insulator, is reached for lower U the higher $J/U$. After $J/U\sim0.2$ however the behaviour becomes more articulated, the instability zone expands in all directions until becoming, for the maximum physical value of $J/U=1/3$ a wide squarish area between $n=2$ and $n\sim2.4$ at all $U>U_c=1.757D$.

\begin{figure}[h!]
\begin{center}
\includegraphics[width=8cm]{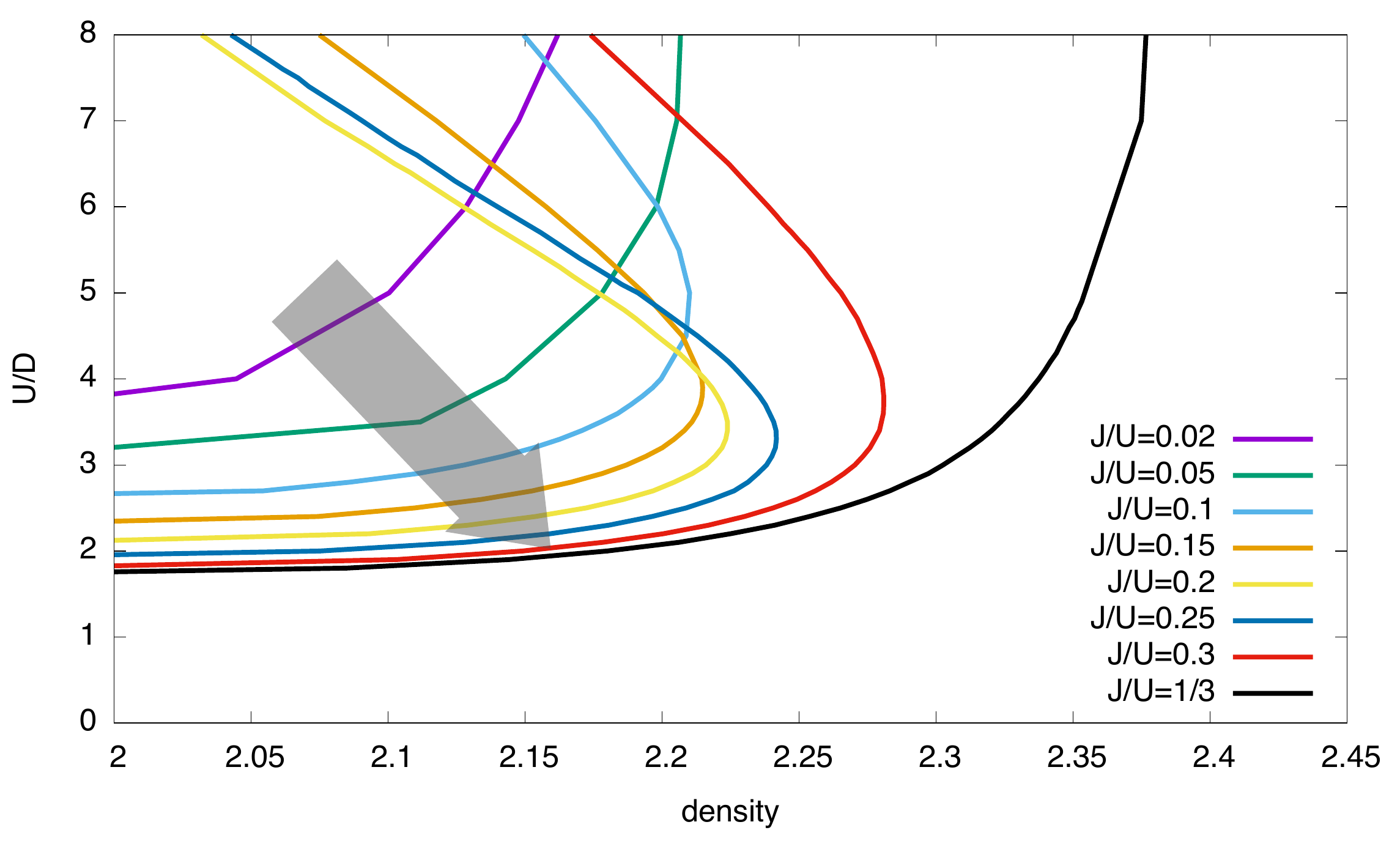}
\caption{Evolution with $J/U$ of the zone of instability (in the U-density plane) in the 2-orbital Hubbard model. At each value the system is unstable for densities between $n=2$ and the corresponding frontier. The gray arrow is a guide to the eye that schematically indicates the effect of an increasing $J/U$. This instability is absent at $J=0$.}
\label{fig:PS_vs_ratioJU}
\end{center}
\end{figure}

\section{Results for the 5-orbital Hubbard model}
\begin{figure}[h!]
\begin{center}
\includegraphics[width=8cm]{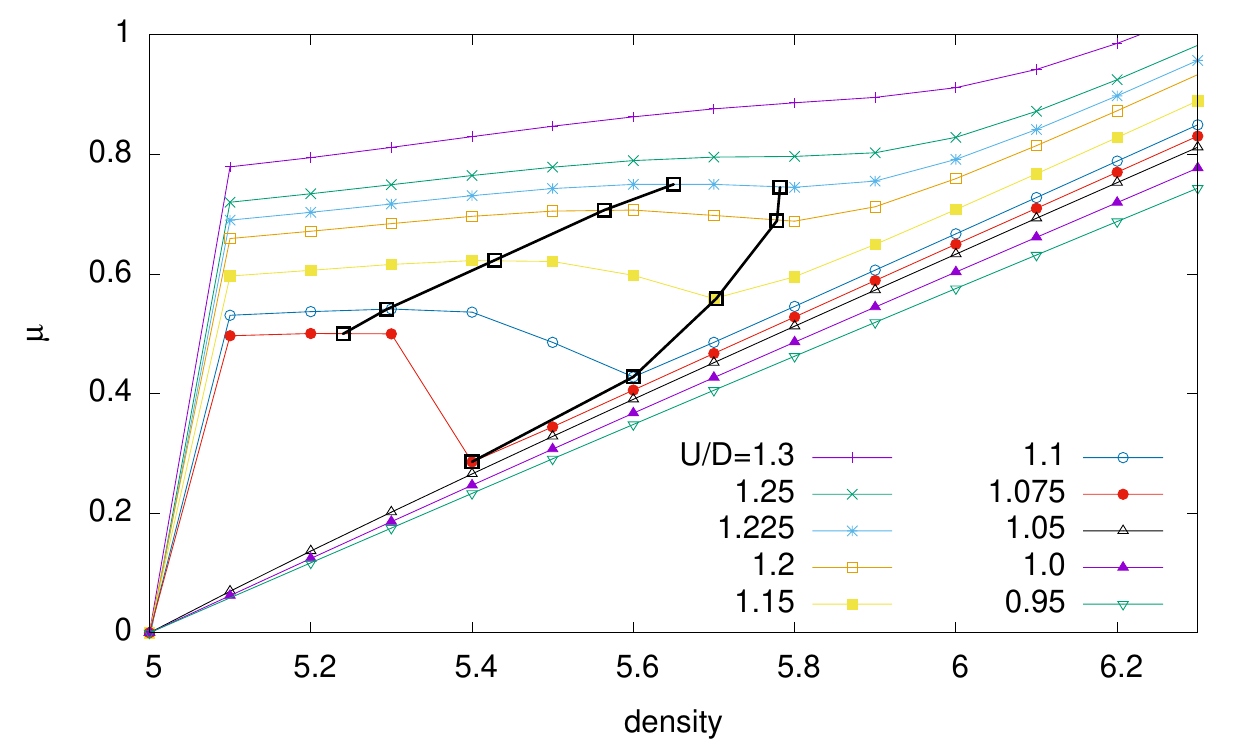}
\caption{Phase separation zone in the 5-orbital degenerate Hubbard models at $J/U=0.25$. Most of the $\mu$ vs $n$ curves show a double change of slope in these models, so that the instability zone extends between two spinodal lines both at finite doping. For growing orbitals the unstable zone extends to larger and larger doping, albeit narrowing in U.}
\label{fig:5orb}
\end{center}
\end{figure}

In Fig. \ref{fig:5orb} the calculation of the $\mu$ vs $n$ curves is reported, for the 5-orbital Hubbard model. As in the 3-orbital case (Fig. 1 in the main article, right panel) a double change of slope is visible in most curves, meaning that the instability happens in a finite range of non-zero doping. This feature, absent in the 2-orbital case, seems to be general for the $M>2$ case. 

The phase diagram has thus a very elongated "moustache" shape in the U-density plane, as visible in the left panel of Fig. 2 in the main article.

\section{Rotiationally invariant slave-boson study of the 2-orbital Hubbard model with spin-flip and pair-hopping interaction terms: confirmation of the phase separation instability.}

It is natural to ask oneself if this unstable zone is a robust physical feature. In particular it is worth checking if the instability is an artifact of the approximate treatment of Hund's coupling due to the chosen density-density Hamiltonian. Slave-spins cannot at present properly treat the full Kanamori hamiltonian at the same approximation level of the Ising one\cite{Sdemedici_Vietri} so that we use the rotationally invariant form of the Kotliar-Ruckenstein slave-bosons \cite{SLechermann_RISB}. Thus implicitly we also show that the presented results are not the artifact of the main method chosen throughout this work (the SSMF) either. 

Indeed, in the 2-orbital model, the occurrence of an instability zone due to a diverging compressibility around the half-filled Mott insulator is confirmed.
This zone, absent at $J=0$ quickly widens its doping range with increasing J/U and it is most extended for $J/U\simeq 0.02$, then decreases again. The extent in doping is smaller than that for the Ising Hamiltonian, reaching densities around n=2.1. 
The low-U frontier of the instability zone still departs from the Mott transition, as in the Ising case.
The shape of the instability zone is analogously re-entrant in both this rotationally invariant and the density density case (i.e. the fact that the frontier $\d_c(U)$ starts at $\d=0$ at the critical interaction for the Mott transition $U=U_c$, quickly grows with $U>U_c$, and then decreases again at larger U until vanishing).

\begin{figure}[h!]
\begin{center}
\includegraphics[width=8cm]{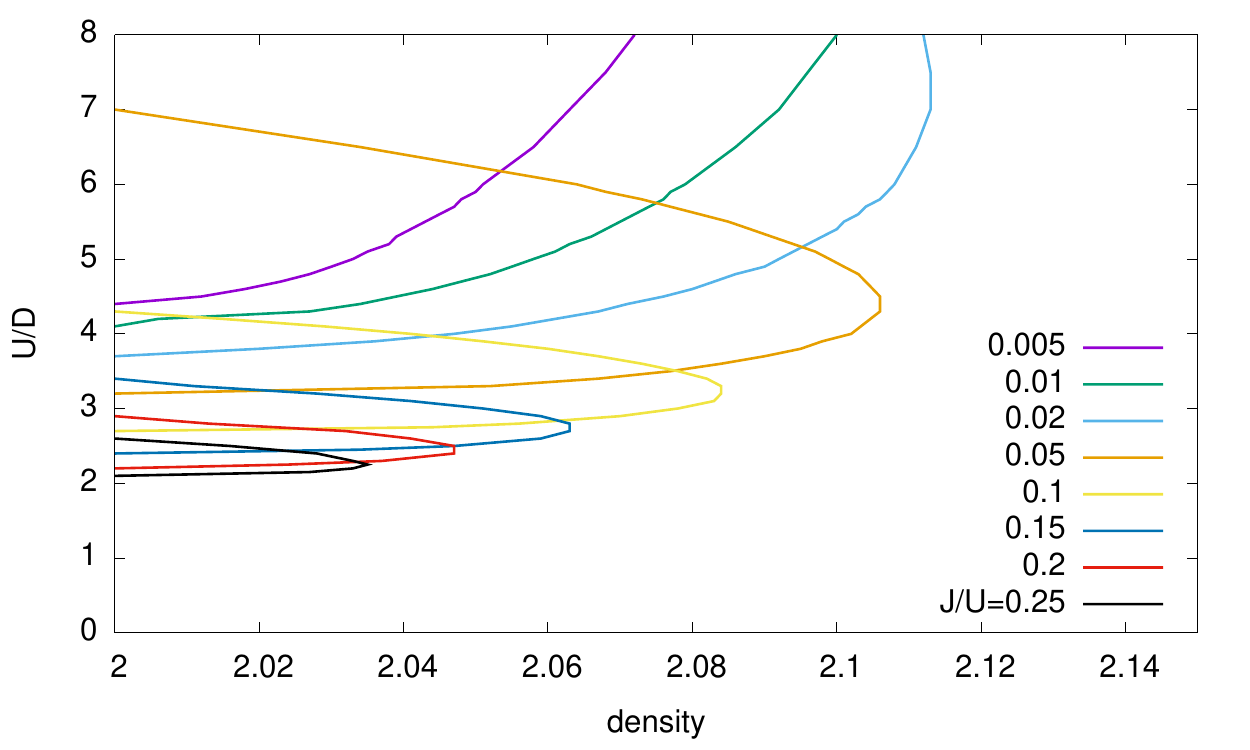}
\caption{Rotationally-invariant slave-bosons study of the 2-orbital Hubbard model with the rotationally-invariant interaction (i.e. $H_{int}$ including the spin-flip and pair-hopping terms, neglected in the rest of the work). Evolution of the zone of phase-separation instability (in the U-density plane) as a function of $J/U$. At each value the system is unstable for densities between $n=2$ and the corresponding frontier.}
\label{fig:PS_vs_ratioJU}
\end{center}
\end{figure}
 Further investigation of the rotationally-invariant case is left for future work, but this preliminary analysis support the robustness of the physical conclusions of the present study.
 
 \section{Proximity of the compressibility enhancement/divergence and the cross-over to the orbitally-decoupled region}
 We have mentioned in the main article that the divergence of the compressibility found in the various studied models departs from the half-filled Mott transition, and coincides (at least close to this transition) with a cross-over, well known in all models and ab-initio studies of systems with strong Hund's coupling\cite{Sdemedici_MottHund,SIshida_Mott_d5_nFL_Fe-SC,Sdemedici_OSM_FeSC,*SYuSi_2-4orbitals,*SYuSi_LDA-SlaveSpins_LaFeAsO,*SLanata_FeSe_LDA+Gutz,*SFanfarillo_Hund}, into a more strongly correlated zone. This zone is characterized by enhanced inter-orbital local ferromagnetic spin-spin correlations (i.e. high-spin configurations dominate in the metallic state), suppressed inter-orbital local charge-charge correlations (i.e. charge excitation in different orbitals are decoupled). This last feature causes (when the symmetry between the different orbitals is broken in the hamiltonian, by e.g. crystal-field, or different hopping) an enhanced selectivity in the correlation strength, where electrons with some orbital character can be strongly correlated while electrons with some other orbital character are weekly correlated. This zone can be labeled with the successful buzzword "Hund's metal"\cite{SYin_kinetic_frustration_allFeSC}, even if a precise commonly accepted definition of what a Hund's metal is not available at the moment.
 
Moving to large dopings the cross-over becomes less sharp than close to half-filling, and the position of the compressibility enhancement seems not to exactly coincide with it. Rather the peak in the compressibility seems to be placed immediately inside the Hund's metal zone, still very close to the cross-over.
\begin{figure}[h!]
\begin{center}
\includegraphics[width=4cm]{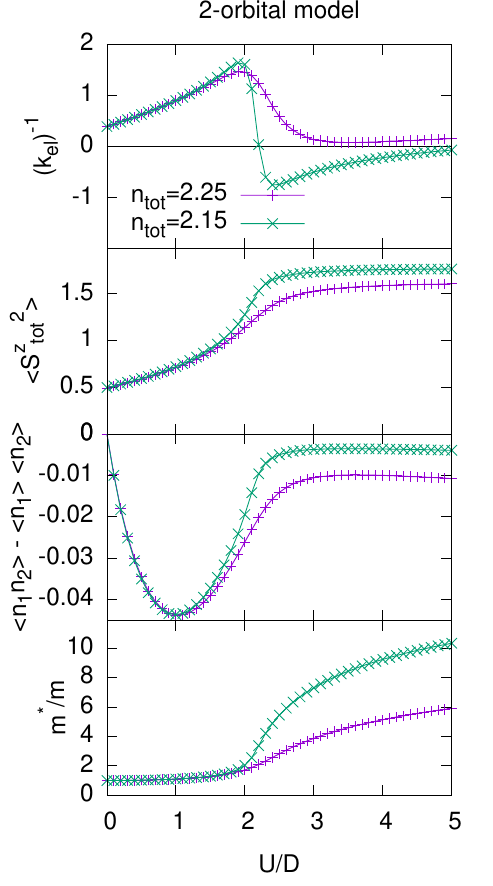}
\includegraphics[width=4cm]{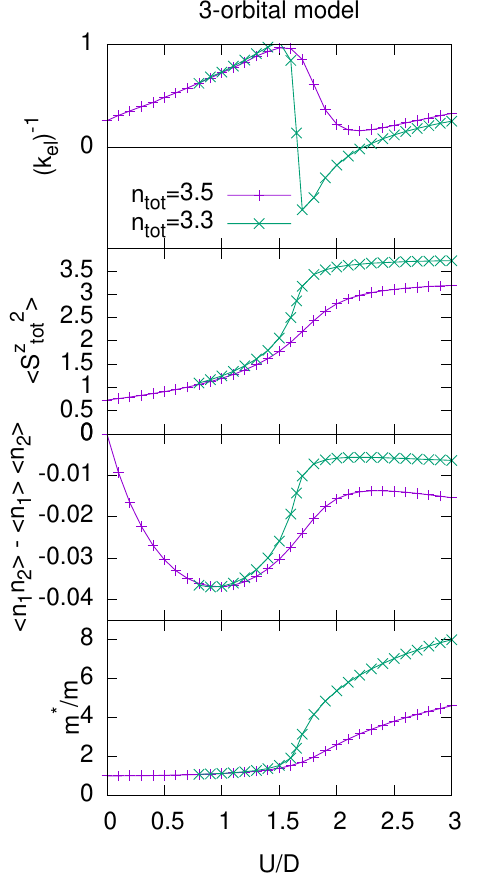}
\caption{Proximity of the compressibility divergence/enhancement with the cross-over into the orbitally-decoupled (that might be identified with the "Hund's metal") phase: 2-orbital (left panel) and 3-orbital (right panel) Hubbard models with Hund's coupling $J/U=0.25$. Results for two densities are reported for each model, as a function of the interaction strength $U/D$: one crossing the instability zone (the smaller density in both cases) and one just outside it (larger density - refer to the phase diagram in the main article, left panel of Fig. 2). Top panels: inverse compressibility showing either a divergence (zero crossing) or an enhancement (minimum) of the compressibility. Upper-middel panels: average of the total z-component of the spin on a given site, showing the entrance in the high-spin ("Hund's") metal. Lower-middle panels: inter-orbital charge-charge correlations, the suppression of which signals the "orbital-decoupling", and independent Mott physics. Bottom panel: the mass enhancement, signalling the stronger correlations of the Hund's metal phase. }
\label{fig:cross-over-peak_models}
\end{center}
\end{figure}

\begin{figure}[h!]
\begin{center}
\includegraphics[width=4cm]{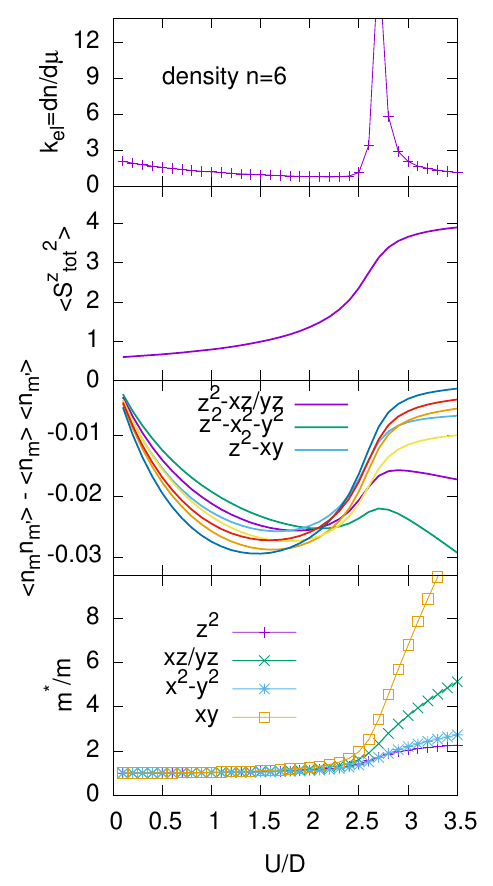}
\includegraphics[width=4cm]{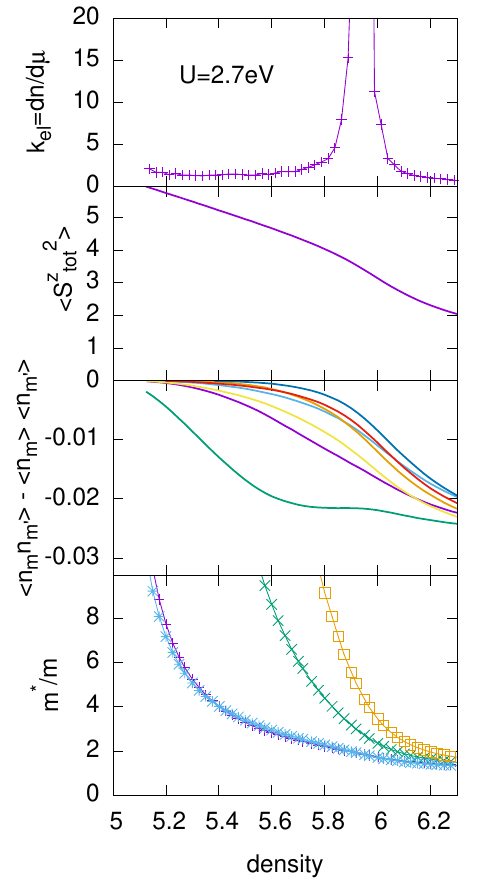}
\caption{Ab-initio DFT+SSMF study of BaFe$_2$As$_2$, additional data to Fig. 3 of the main article. Top panel: compressibility, remaining panels: same quantities than in Fig. \ref{fig:cross-over-peak_models}. In this realistic simulation the proximity of the compressibility enhancement to the cross-over into the orbitally-decoupled "Hund's metal" phase is even sharper than in the model studies.}
\label{fig:cross-over-peak_BaFe2As2}
\end{center}
\end{figure}

The results illustrating the coincidence/proximity of the frontier to the Hund's metal and the compressibility divergence/peak are reported for the 2-orbital and the 3-orbital models in Fig. \ref{fig:cross-over-peak_models} and for the realistic case of BaFe$_2$As$_2$ in Fig. \ref{fig:cross-over-peak_BaFe2As2}.
 
\section{Formulas for the compressibility from the renormalized quasiparticle Hamiltonian}

For the system described by the Hamiltonian 
\be\
H-\mu N= \sum_{km\s} (Z\epsk+\lambda-\mu) f^\+_{km\s}f_{km\s},
\ee
the total number of quasiparticles, equal to the particle density, reads
\be
n_f=\int^{(\mu-\lambda)/Z} d\eps D(\eps)=n.
\ee
The electronic compressibility $\kappa_{el}=\frac{d n}{d \mu}$ is obtained by differentiation:
\be
\frac{d n}{d \mu}=D(\frac{\mu-\lambda}{Z})\frac{1}{Z}\left(1-\left[(\frac{\mu-\lambda}{Z})\frac{dZ}{d\mu}+\frac{d\lambda}{d\mu}\right]\right),
\ee
and then, using the simple identity $\frac{dX}{d\mu}=\frac{dX}{dn} \frac{dn}{d\mu}$ this yields:

\begin{align}
\frac{d n}{d \mu}
=\frac{D(\frac{\mu-\lambda}{Z})/Z}{1+\frac{D(\frac{\mu-\lambda}{Z})}{Z}[(\frac{\mu-\lambda}{Z})\frac{dZ}{dn}+\frac{d\lambda}{dn}]}
=\frac{D(\mu_0)/Z}{1+\frac{D(\mu_0)}{Z}[\mu_0\frac{dZ}{dn}+\frac{d\lambda}{dn}]}\nonumber \\
=\frac{D^*(\mu)}{1+D^*(\mu)[\mu_0\frac{dZ}{dn}+\frac{d\lambda}{dn}]}
=\frac{1}{D^*(\mu)^{-1}+[\mu_0\frac{dZ}{dn}+\frac{d\lambda}{dn}]}.
\end{align}
Here we have used the fact that the Luttinger theorem imposes that $\mu_0=\frac{\mu-\lambda}{Z}$, where $\mu_0$ is the chemical potential for the non-interacting system with the same particle density, and the fact that $D^*(\mu)=D(\mu_0)/Z$. 
Indeed the Luttinger theorem states that the volume of the interacting Fermi surface is proportional to the particle density. i.e. $n_f=n$, so that this useful expression can be found, relating the chemical potential in our interacting model to that of the non-interacting case with the same density. Indeed by equating the densities of particles at U=0 with the interacting quasiparticle density at zero temperature i.e. 
\be
n(U=0)=\int^{\mu_0} d\eps D(\eps)=\int^{(\mu-\lambda)/Z} d\eps D(\eps)=n_f(U),
\ee one obtains $\mu_0=\frac{\mu-\lambda}{Z}$.

Thus by comparing to the Fermi-liquid formulas 
\be\label{kappa_FL}
\kappa_{el}=\frac{D^*(\eps_F)}{1+F_0^s}=\frac{1}{D^*(\eps_F)^{-1}+f_0^s},
\ee 
we can identify
\be\label{F0s}
F_0^s=D^*(\mu)\left[\mu_0\frac{dZ}{dn}+\frac{d\lambda}{dn}\right], \qquad f_0^s=\mu_0\frac{dZ}{dn}+\frac{d\lambda}{dn}.
\ee

\begin{figure}[h!]
\begin{center}
\includegraphics[width=9cm]{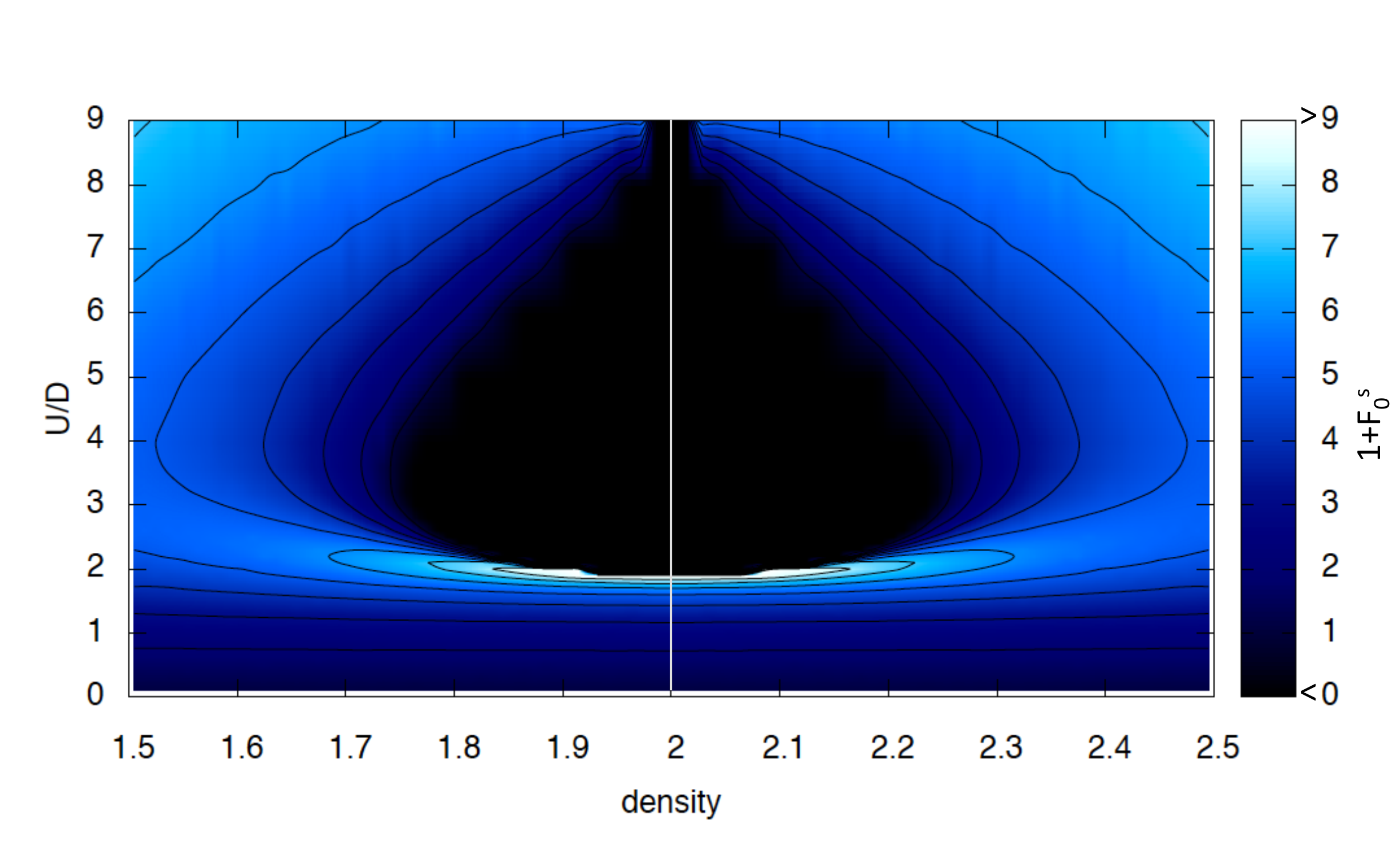}
\caption{Map of the Landau Fermi-liquid parameter $1+F_0^s$ for the 2-orbital model ith $J/U=0.25$. The black zone is where this quantity is negative and the Fermi liquid is correspondingly unstable.}
\label{fig:Hubbard_bands}
\end{center}
\end{figure}

\section{Role of the quasiparticle energy shift $\lambda$ in the enhancement/divergence of compressibility}

Following formulas (\ref{kappa_FL}) and (\ref{F0s}), the compressibility diverges, if $D^*(\mu)$ remains finite (i.e. the quasiparticle mass does not diverge), when $F_0^s=-1$. This implies that $D^*(\mu)^{-1}+\mu_0\frac{dZ}{dn}+d\lambda/dn$ becomes negative by going through zero. 
\begin{figure}[h!]
\begin{center}
\includegraphics[width=8cm]{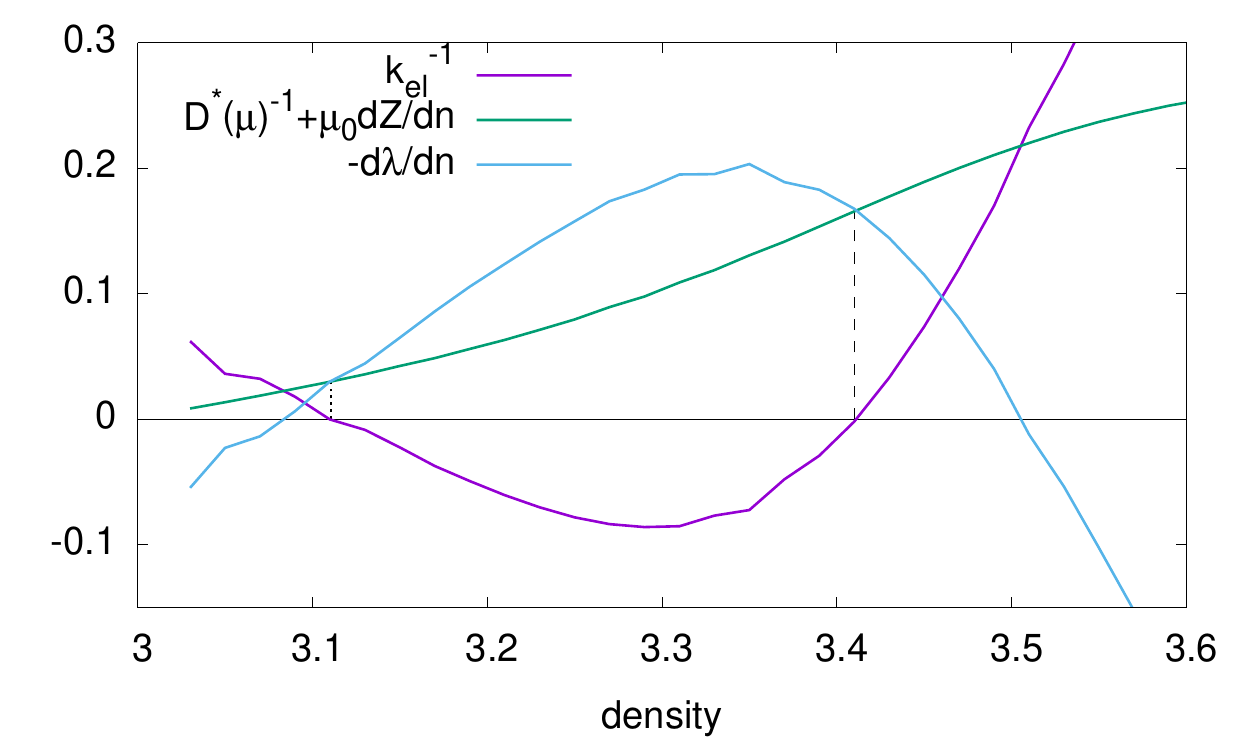}
\caption{Example illustrating the role of the quasiparticle energy shift $\lambda$ in determining the divergence of the compressibility. Here we report the result of the SSMF for the 3-orbital Hubbard model with $U/D=2.1$ and $J/U=0.25$. The dashed lines show that the compressibility diverges when $-d\lambda/dn$ reaches the value of $D^*(\mu)^{-1}+\mu_0\frac{dZ}{dn}$ which is a positive quantity. Thus $d\lambda/dn$ has to be negative.}
\label{fig:dlamdn_neg}
\end{center}
\end{figure}
However $D^*(\mu)^{-1}+\mu_0\frac{dZ}{dn}>0$, because the quasiparticle DOS is a definite positive quantity and $\frac{dZ}{dn}$ has always the sign of $\mu_0$ near the half-filled Mott insulator in our particle-hole symmetric models. This is because $\mu_0=0$ at half-filling in these cases (thus $\mu_0$ has the sign of the doping from half-filling), and Z decreases monotonically when approaching the Mott insulating state (thus $\frac{dZ}{dn}$ has also the sign of the doping). 

This entails that $d\lambda/dn$ has to become negative, and this is what is found numerically in all the examined cases in this article, as shown in Fig. \ref{fig:dlamdn_neg}.

\end{document}